\documentclass[11pt,twoside]{article}

\usepackage{asp2006}
\usepackage{epsf}
\usepackage{psfig}
\usepackage{lscape}

\begin{document}
\title{Constraints on the prevalence of luminous, dusty starbursts
at very high redshifts}
\author{R.\ J.\ Ivison}
\affil{Royal Observatory, Blackford Hill, Edinburgh, Scotland}

\begin{abstract} At a conference devoted to ever deeper surveys
hunting for ever more distant galaxies, I posed a question for which a
concensus view has been difficult to reach: `Is there evidence for, or
can we rule out, a significant population of dust-obscured starbursts
at $z\rm > 3$?'  If, as seems likely, submm-selected galaxies are
proto-ellipticals, one of the biggest unanswered questions is whether
a significant fraction form at very high redshift -- perhaps by the
collapse of single gas clouds -- or whether the entire population
forms over a range of redshifts, primarily via merging. The latter
scenario is favoured strongly by existing data for the majority of
bright submm galaxies -- mergers are common; typical spectroscopic
redshifts are in the range 1--3. However, our reliance on radio
imaging to pinpoint submm galaxies leaves open the possibility of a
significant population of very distant, massive starbursts. To rule
out such a scenario requires a completely unbiased redshift
distribution for submm-selected galaxies and this is unlikely to be
forthcoming using conventional optical/infrared spectroscopic
techniques. Here, I summarise recent attempts to close that door, or
pass through to an early Universe inhabited by a significant
population of collosal, dust-obscured starbursts. I conclude that the
door is only barely ajar; however, the idea of galaxies forming, near
instantaneously, on a collosal scale is not dead: I show an SMG imaged
by MERLIN+VLA for $>$1\,Ms -- the deepest high-resolution radio image
so far obtained. The data reveal a galaxy-wide starburst covering
$\ge$10\,kpc. Thus interpreting SMGs in terms of compact ULIRG-like
events may not always be appropriate, as one might expect when
considering the Eddington limit for $\ge$10$^3$-M$_{\odot}$\,yr$^{-1}$
starbursts.
\end{abstract}

\section{Introduction}

\setcounter{figure}{0}
\begin{figure}[!t]
\plottwo{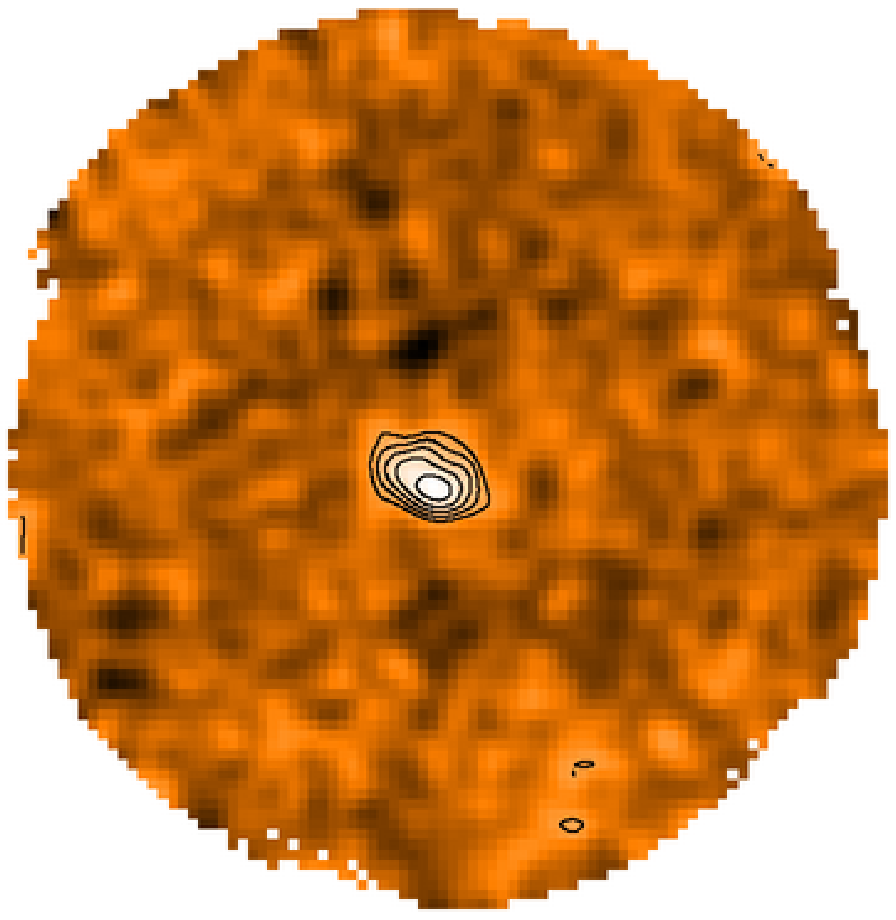}{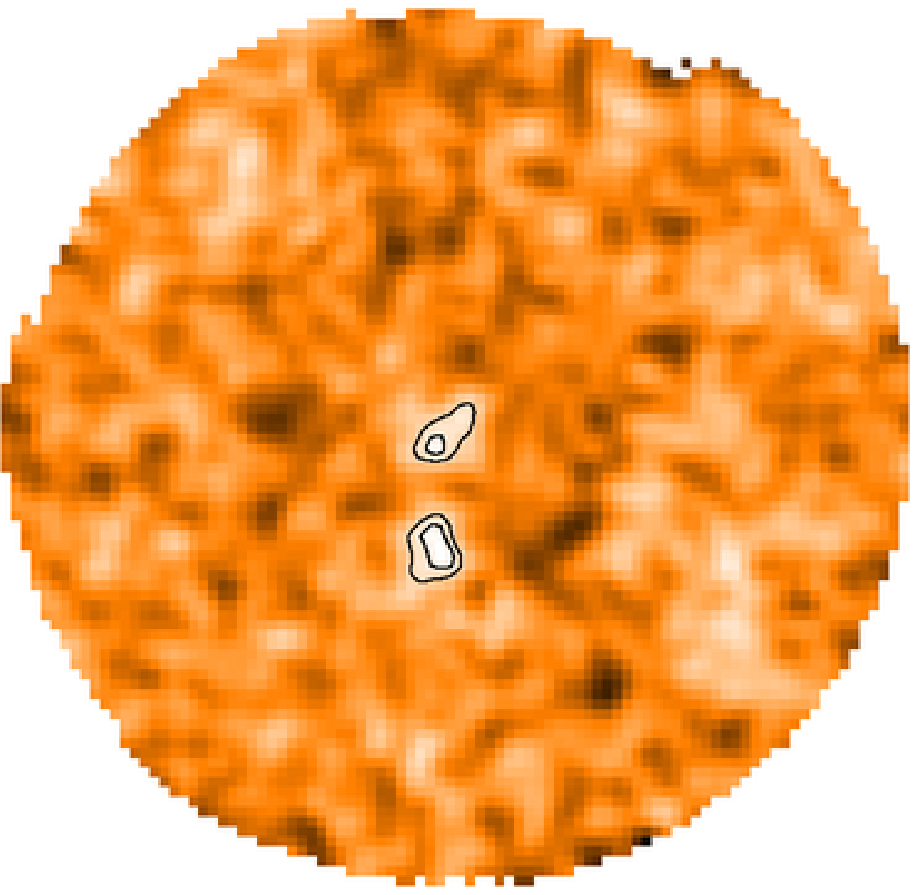}
\caption{850-$\mu$m images of distant, optically-selected
quasars (Priddey, Ivison \& Isaak, in preparation). {\itshape Left:\/}
SDSS\,J075618.13+410408.5 at $z=\rm 5.1$; {\itshape right:\/}
SDSS\,J104433.04-012502.2 at $z=\rm 5.8$, with a nearby SMG that may
lie at the same redshift (see also Ivison et al.\ 2000; Stevens et
al.\ 2003).}
\end{figure}

\setcounter{figure}{1}
\begin{figure}[!t]
\plotfiddle{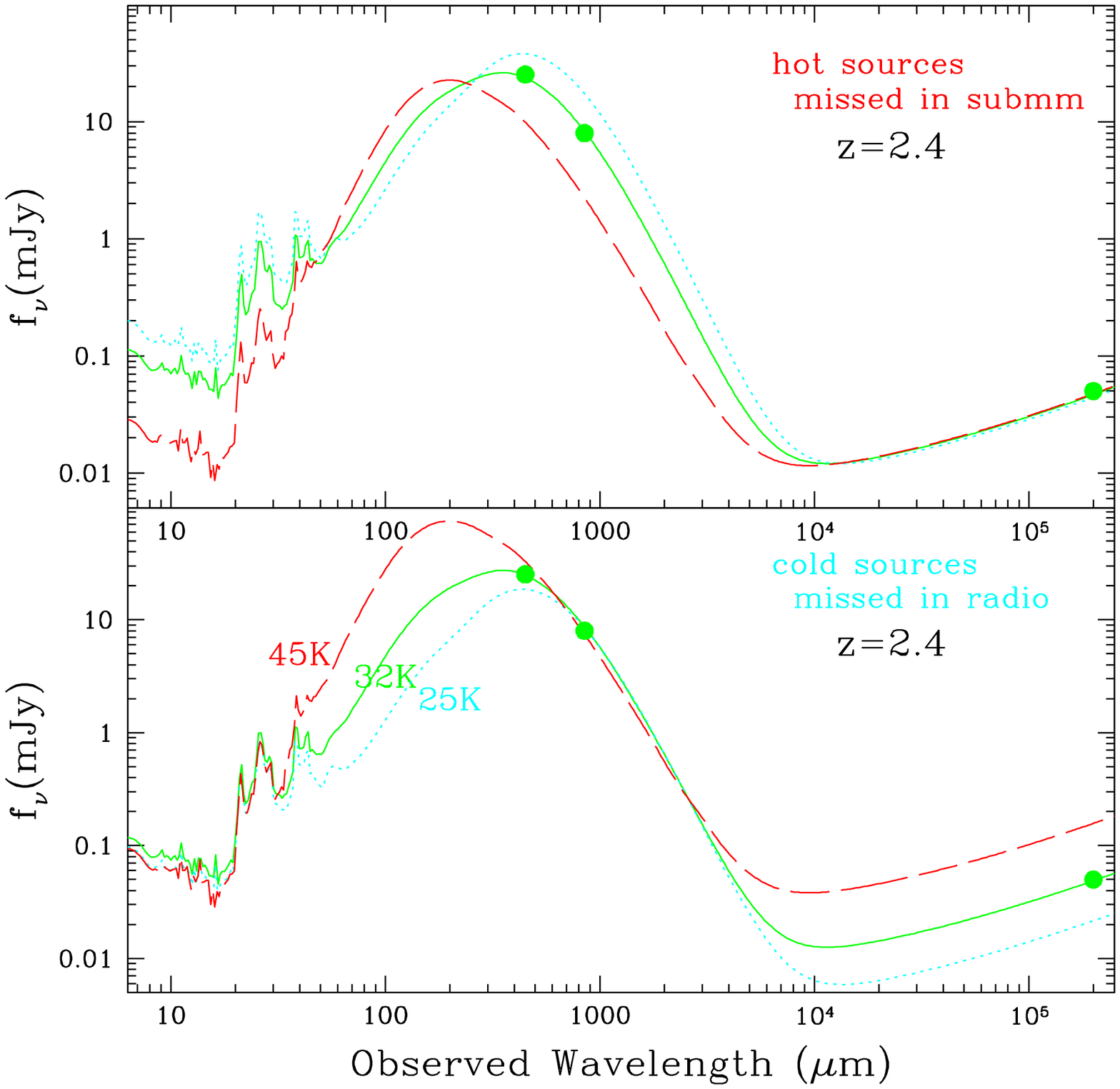}{7.5cm}{0}{43}{43}{-120}{-75}
\caption{Possible spectral energy distributions (SEDs) describing the
emission from a typical SMG at $z=\rm 2.4$, with 850-$\mu$m and
1.4-GHz flux densities of 6\,mJy and 50\,$\mu$Jy, respectively, taken
from C05.  Overlaid are SEDs for $T_{\rm dust}$ = 25, 32 and
45\,K. {\itshape Upper panel:\/} SEDs normalised to the radio datum,
emphasising how galaxies with hotter dust (and lower implied dust
masses) are missed in the submm at $z\ge\rm 2$. {\itshape Lower
panel:\/} SEDs normalised at 850\,$\mu$m, highlighting how sources
with cooler dust are undetectable in the radio at high redshifts. The
need to pinpoint the counterpart of an SMG means that the coldest
SMGs, which will lie below typical radio flux limits, will be missed
by spectroscopic surveys. This bias becomes potentially serious at the
highest redshifts where even relatively warm SMGs fall beneath typical
radio limits.}
\end{figure}

Surveys with bolometer arrays at mm and submm wavelengths are
potentially sensitive to dusty objects at extreme redshifts, galaxies
that drop out of surveys at shorter and longer wavelengths due to
obscuration and unfavourable $K$ corrections. The first cosmological
surveys, using the SCUBA and MAMBO cameras, radically changed the
accepted picture of galaxy formation and evolution, moving away from
the optocentric view of the last century. The discovery of so-called
`SCUBA galaxies' (Smail, Ivison \& Blain 1997), now better known as
submm galaxies or SMGs, was greeted with surprise due to the
remarkable evolution in dusty, starburst galaxies implied by such a
large source density at the flux levels accessible to the first
generation of bolometer arrays.

Sadly, SCUBA has been retired. SCUBA-2 will revolutionise submm
astronomy once again, but not until the winter of 2007-08.  At the
depths that were accessible to SCUBA, and to cameras like MAMBO-117,
SMGs constitute around half of the submm background and thus at least
a quarter of the total extragalactic background. The relation between
metal production and background radiation then implies that roughly
25\% of all the stars ever formed have formed in extreme objects like
this (Eales et al.\ 1999). The combination of this argument, their
space density and their high bolometric luminosities made it hard to
resist the conclusion that these objects are ellipticals seen during a
phase in which a significant fraction of their stars are forming,
loosely `proto-ellipticals'.

Initial excitement was tempered by the first efforts to study SMGs at
optical and infrared wavelengths. Early reports, backed up with a
study in the GOODS North field by Hughes et al.\ (1998), suggested
that the majority of the submm population had no plausible 
counterparts at magnitudes routinely accessible to optical
surveys. Attention was diverted to various redshift engines and
broadband photometric techniques (e.g.\ Townsend et al.\ 2001; Wiklind
2003). As a result, only a handful of detailed studies were attempted,
often for extreme and possibly unrepresentative galaxies.

Progress resumed when deep, wide-field, high-resolution radio imaging
of submm survey fields were shown to recover roughly half of the SMGs
in each field, with an astrometric precision of $\sim$0.3$''$ (e.g.\
Ivison et al.\ 1998, 2002). Combined with the submm flux density,
radio data provide a rough estimate of redshift (Carilli \& Yun
1999). They also enable refinement of submm samples, increasing the
detection fraction to two thirds of SMGs at 850-$\mu$m flux density
levels in excess of $\sim$5\,mJy (Ivison et al.\ 2002; Pope et al.\
2006). Accurate positions in hand, bright SMGs were found to be a
diverse population --- some are quasar-like, with broad lines and X-ray
emission (e.g.\ Knudsen et al.\ 2003), some resemble BAL quasars
(e.g.\ Vernet \& Cimatti 2001), some are morphologically complex (e.g.\
Smail, Smith \& Ivison 2005), some are extremely red (e.g.\ Smail et al.\
1999; Webb et al.\ 2003; Dunlop et al.\ 2004); some
bear the signatures of obscured active nuclei and/or
powerful winds (e.g.\ Smail et al.\ 2003).

\setcounter{figure}{2}
\begin{figure}[!t]
\plotfiddle{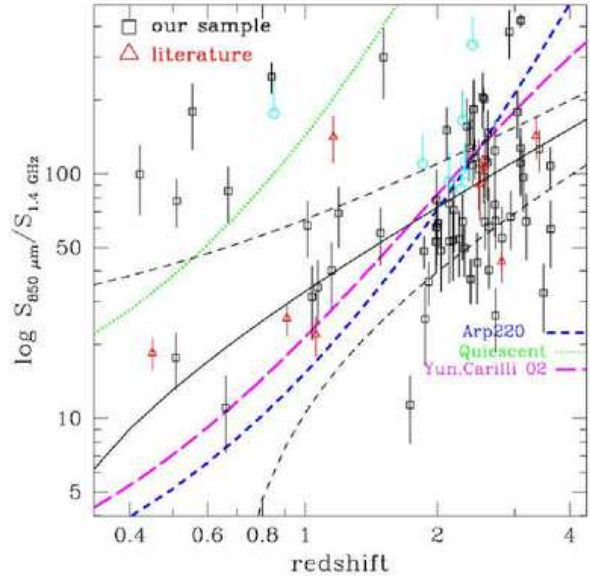}{7cm}{0}{80}{80}{-110}{-320}
\caption{The 850-$\mu$m/1.4-GHz flux density ratio versus redshift for
the radio-detected SMGs with spectroscopic redshifts, taken from
C05. The predicted variation in flux density ratio for three SEDs are
shown, including that of Arp\,220. If the SMG population displayed a
narrow range of SEDs we would see them tracing just such a track in
this plot. The trend of submm/radio flux density ratio with redshift
is flatter than the template tracks. The wide scatter, even at fixed
redshift, suggests a wide variety of SEDs. The shaded region shows the
$\pm\sigma$ envelope of the dispersion calculated in three redshift
bins containing equal numbers of SMGs. The radio selection criterion
for existing spectroscopic work biases those surveys to galaxies with
enhanced radio emission, such as Mrk\,231.  A purely submm-selected
sample would show an even wider range in submm/radio flux density
ratio than the one shown here. The range of SMG SEDs
means that simple photometric redshift techniques struggle to
predict accurate redshifts for individual galaxies.}
\end{figure}

Not surprisingly, given their median magnitude ($R=\rm 25$, Vega),
spectroscopic redshifts have been difficult to determine. The most
ambitious survey of SMGs was undertaken by Chapman et al.\ (2003, 2005
--- hereafter C03, C05). The median redshift was $\sim$2.2 for $S_{\rm
850\mu m} \ge \rm 5$-mJy galaxies selected using SCUBA and pinpointed
at 1.4\,GHz or, in some cases, selected as optically faint radio
galaxies and detected subsequently with SCUBA. The accurate redshifts
reported by C03 and C05 facilitated the first systematic measurements
of molecular gas mass for SMGs ($\rm 10^{10}$--$\rm
10^{11}\,M_{\odot}$) via observations of CO (Neri et al.\ 2003; Greve
et al.\ 2005; see also early work by Frayer et al.\ 1998, 1999), as
well as constraints on gas reservoir size and dynamical mass (Tacconi
et al.\ 2006 -- though see later). The data suggest SMGs are massive
systems and provide some of the strongest tests of galaxy-formation
models to date.

\section{Current state of play}

In spite of this progress, a detailed understanding of SMGs remains a
distant goal that many believe is now best achieved by detailed study
of individual objects. Confusion currently limits our investigations
to the brightest SMGs (although surveys through lensing clusters have
provided a handful of sources typical of the faint population that
dominates the cosmic background, e.g.\ Smail et al.\ 2002; Kneib et
al.\ 2004; Borys et al.\ 2004). We must also recall that selection
biases have potentially skewed our understanding. Around half of all
known SMGs remain undetected in the radio due partly, at least, to the
lack of sufficiently deep radio data, which do not benefit from the
same $K$ correction as submm data. Although a handful of
radio-undetected SMGs have been targeted spectroscopically, few have
yielded reliable redshifts. Finally, there has been only limited
coverage at red and IR wavelengths in existing spectroscopic
surveys.

Here, at a conference devoted to ever deeper surveys hunting for ever
more distant galaxies, I posed the question: ``Is there evidence for,
or can we rule out, a significant population of dust-obscured
starbursts at $z\rm > 3$?''

It is difficult to present a concensus view: this issue divides the
submm community, with opinions correlated distinctly with age. Deep
imaging was not possible in the pre-SCUBA era so submm cosmologists
spent their time hunting for thermal emission from distant quasars and
radio galaxies (Fig.~1; e.g.\ McMahon et al.\ 1994; Ivison 1995;
Hughes, Rawlings \& Dunlop 1997). In the eyes of the SCUBA generation,
these saggy, old astronomers have found it difficult to accept that
their dusty high-redshift AGN were rare freaks which modellers can
label `unrepresentative'. The C05 spectroscopic survey suggests that
the vast majority of SMGs lie at $z\rm = 1-3$ with a mere smattering
at $z\rm <1$ and $z\rm >4$: 20\% at most (cf.\ Dunlop 2001). But is
this necessarily so? Would we be aware of a significant population of
distant, dusty starbursts, maybe at the $\ge$25\% level, hiding within
current SMG samples?

Several potentially important effects -- not least our reliance on
radio imaging to pinpoint SMGs (see Fig.~2) -- leave the door open for
a significant population of distant, massive starbursts. To rule out
such a scenario requires an unbiased redshift distribution for SMGs
and this is unlikely to be forthcoming by conventional spectroscopic
techniques (though see Pope et al.\ 2006 for the best available
attempt using conventional optical/IR photometric methods).  Before
ALMA, {\em JWST} or the ELT, and without the construction of dedicated
wide-bandwidth cm/mm/submm spectrometers designed to solve this
specific problem, the only way to make progress is to use the
distinctive SEDs of high-redshift dusty galaxies as a method of
estimating redshifts.

\section{Recent work}

\setcounter{figure}{3}
\begin{figure}[!t]
\plotfiddle{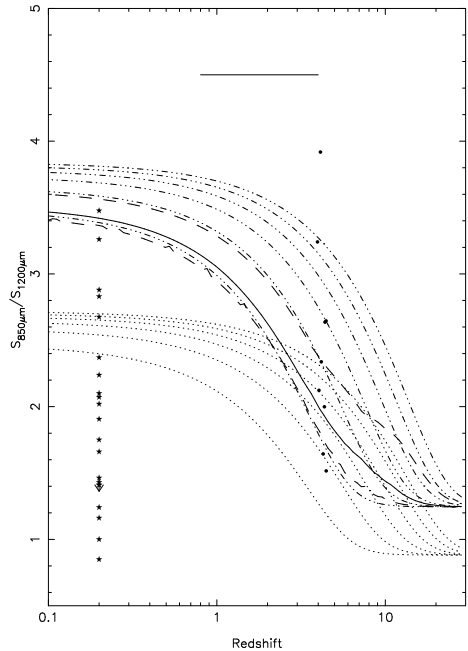}{8cm}{0}{35}{35}{-100}{-20}
\caption{Predicted 850-/1200-$\mu$m flux ratio as
a function of redshift from Eales et al.\ (2003). Asterisks show
measured values of the flux ratio for MAMBO-selected sources, plotted
arbitrarily at $z=\rm 0.2$; dots: high-redshift
quasars; horizontal line: range of spectroscopic redshifts for SMGs,
circa 2003; thin lines: flux ratio versus redshift predicted
for single-temperature dust models; dotted lines: $\rm \beta = 1$;
dot-dashed lines: $\rm \beta = 2$. For both sets, the lowest and
highest lines are for $T_{\rm dust}$ = 20 and 70\,K, respectively,
with other lines at intervals of 10\,K. Thick lines: predictions based
on the models of Dunne \& Eales (2001). Continuous line: median
predicted value of the flux ratio; dashed lines: lowest and highest
predicted value at each redshift.}
\end{figure}

\setcounter{figure}{4}
\begin{figure}[!t]
\plotfiddle{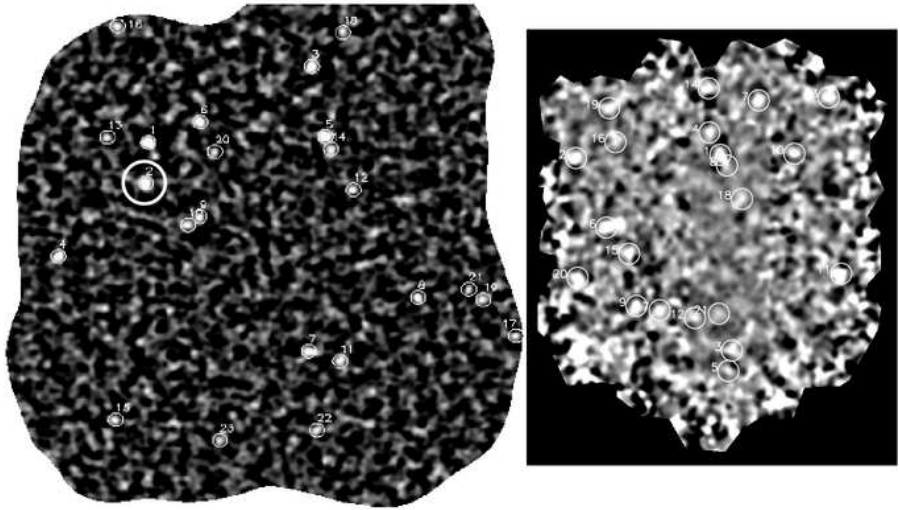}{6.8cm}{0}{80}{80}{-180}{-310}
\caption{{\itshape Left:\/} MAMBO 1200-$\mu$m image of the Lockman
Hole East field from Greve et al.\ (2004). Sources detected at
$\ge$$3.5\sigma$ are circled. The large white circle indicates the
brightest MAMBO source that lacks a counterpart in the SCUBA
850-$\mu$m image of the same region: LH\,1200.02, the second
most significant source in the MAMBO image -- see also Fig.~6.
The 850-$\mu$m image is shown, at roughly the same scale, to the
{\itshape right}, this time smoothed with a beam-size Gaussian
(14.5$''$ {\sc fwhm}), from Scott et al.\ (2002).}
\end{figure}

\setcounter{figure}{5}
\begin{figure}[!t]
\plotfiddle{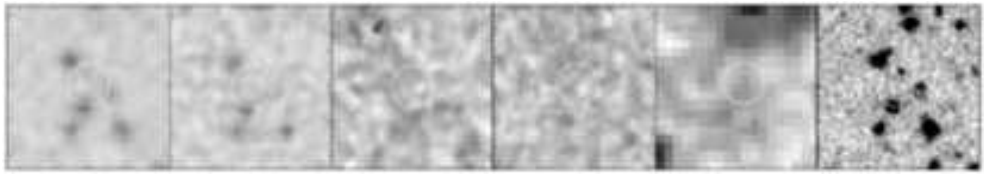}{1.5cm}{0}{126}{126}{-185}{-915}
\caption{{\itshape Spitzer} imaging at 3.6, 4.5, 5.8, 8 and 24\,$\mu$m
of the `SCUBA drop-out', LH\,1200.02 (courtesy: Eiichi Egami). The
final panel shows a deep $R$-band image. The counterpart to the SMG is
far from obvious.}
\end{figure}

\setcounter{figure}{6}
\begin{figure}[!t]
\plotfiddle{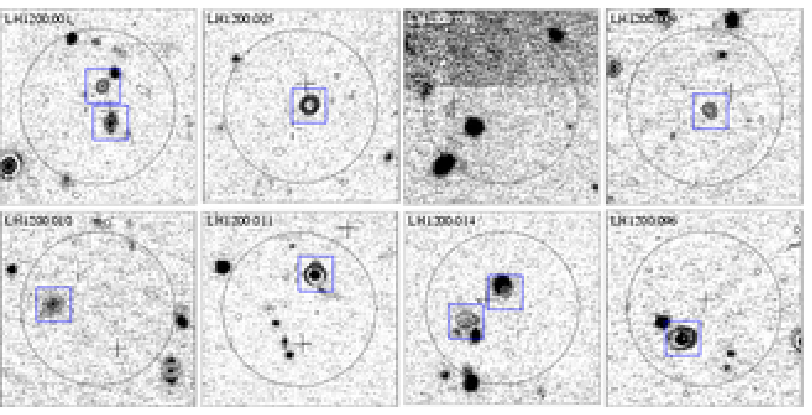}{5.8cm}{0}{150}{150}{-175}{-955}
\caption{Postage stamps of galaxies from the dual-survey sample of
Ivison et al.\ (2005). Extremely deep near-IR data are shown as a
greyscale ($K=\rm 20.9$, 3$\sigma$, Vega), upon which 1.4-GHz contours
are plotted.  Open crosses: SCUBA positions; solid crosses: MAMBO
positions; 8-arcsec-radius circles: average positions; boxes mark
robust radio sources. A remarkable number of SMGs have counterparts
that are barely visible at $K\sim\rm 21$ (Vega). The median optical magnitude
of this SMG sample is $R=\rm 25.0$ (Vega).}  \end{figure}

Carilli \& Yun (1999) and Eales et al.\ (2003) were amongst the first
to address this issue, along with Hughes et al.\ (2002), Wiklind
(2003) and Aretxaga et al.\ (2003), though cf.\ Blain, Barnard \&
Chapman (2003) and C05. The approach generally employs long-wavelength
flux ratios as redshift indicators. Carilli \& Yun pioneered the use
of the $S_{\rm 850\mu m}/S_{\rm 1.4GHz}$ flux density ratio (Fig.~3),
mentioned first by Hughes et al.\ (1998) and used later by Ivison et
al.\ (2002) to determine a median redshift for SMGs of $\ge$2;
however, this particular ratio flattens at higher redshifts and the
flux ratio is not very useful at $z>\rm 2$.

Eales et al.\ (2003) developed a different redshift indicator: the
ratio of 850- to 1200-$\mu$m flux density, as traced by MAMBO and
SCUBA, respectively. This has complementary properties to the
radio/submm flux ratio: it is useless as a redshift indicator at
$z<\rm 2.5$, but it should very good at higher redshifts. For
answering the specific question of whether there is a significant
population of dusty, starburst galaxies at $z >\rm 3$, as one would
expect if part of the elliptical population forms by monolithic
collapse, this is the most useful redshift estimator.  Fig.~4 shows
the predicted 850-/1200-$\mu$m flux density ratio as a function of
redshift for several template SEDs. Errors for individual redshift
estimates are clearly very large, to put it mildly, but the error on
the mean redshift of a large sample is much smaller.

Eales et al.\ carried out SCUBA 850-$\mu$m photometry-mode
observations (in which a single bolometer in each array is pointed at
the target) of the brightest sources from one of the earliest
blank-field 1200-$\mu$m MAMBO surveys. Using SCUBA to follow up MAMBO
sources exploits two advantages over MAMBO follow-up of SCUBA sources:
first, a 1200-$\mu$m survey is likely to contain a larger fraction of
very-high-redshift sources than an 850-$\mu$m survey; second, the IRAM
30-m MRT's resolution at 1200\,$\mu$m is 30\% better than the SCUBA
resolution at 850\,$\mu$m which is crucial for accurate
photometry. More accurate positions from radio imaging would be better
still, but this would re-introduce our most pernicious bias.

Eales et al.\ concluded that the 850-/1200-$\mu$m flux density ratios
for 15 of their sample of 23 SMGs were much lower than expected for
low-redshift galaxies and that these SMGs must either be at higher
redshifts, $z>\rm 3$ in fact, or have different rest-frame SEDs that
local starbursts.  Various systemic effects were investigated and
ruled out, including concerns about absolute flux calibration,
astrometric errors and flux boosting, the latter being a particular
example of the process described by Eddington (1940) in which the
statistical properties of a a distribution of experimental
measurements are distorted by noise.

Greve et al.\ (2004) took the next logical step, using unbiased
blank-field surveys at 850 and 1200\,$\mu$m (Fig.~5) to investigate
the reliability of SMG samples, to analyse SMGs using flux ratios
sensitive to redshift at $z\rm > 3$, and to search for `SCUBA
drop-outs', i.e.\ $z\rm >> 3$ galaxies, where the far-IR bump has
moved through SCUBA's 850-$\mu$m filter. For a sub-sample of 13 SMGs
detected by both MAMBO and SCUBA, Greve et al.\ concluded that the
distribution of 850-/1200-$\mu$m flux density ratios was consistent
with the spectroscopic redshift distribution of radio-detected SMGs
(C03, C05). For the 18 MAMBO sources not detected by SCUBA, the
distribution of 850-/1200-$\mu$m flux density ratios was
indistinguishable statistically from that of the SMGs identified by
both MAMBO and SCUBA.

Although Greve et al.\ (2004) disagreed fundamentally with the
findings of Eales et al.\ (2003), it is easy to see that the
sub-sample of SMGs detected by both MAMBO and SCUBA -- which makes up
less than half of the parent MAMBO sample -- will (by definition)
contain no `SCUBA drop-outs'. At this stage I therefore consider that
the door to an early Universe inhabited by a significant population of
collosal, dust-obscured starbursts is still ajar. In fact, the sample
contains at least one object detected robustly (6.8$\sigma$) at
1200\,$\mu$m and yet apparently absent in the SCUBA image -- an object
with the very characteristic that Greve et al.\ set out to find. This
SMG (LH\,1200.02 at 10 52 38.8, +57 23 22, J2000) is circled in
Fig.~5.  It lacks both a robust radio detection and a convincing
{\itshape Spitzer} identification (Fig.~6). Elucidating its nature --
hyperluminous, dusty starburst at $z>\rm 3$ or an less distant object
with significantly lower characteristic dust temperature, possibly
$\le$25\,K -- has progressed at an excruciatingly slow pace.

Ivison et al.\ (2005) presented a robust sample of bright SMGs
selected using a dual-survey extraction technique, aiming to leave the
regime where the modest significance of sources compromises the
analyses via the presence of spurious sources or the effect of flux
boosting. We would expect the resulting sample to be detected in
well-matched radio imaging ($\sigma_{\rm 850\mu m}/\sigma_{\rm 1.4GHz}
\sim\rm 500$). The goal was to determine the true fraction of radio
drop-outs amongst SMGs, as well as some practical information such as
the intrinsic positional uncertainty for SMGs in the absence of
radio/IR counterparts.

The dual-survey selection technique yielded 19 SMGs of which a high
fraction ($\sim$80\%) had radio counterparts (Fig.~7). Given the
likely $\sim$10\% contamination by spurious sources, this suggests
that very distant SMGs ($z\gg\rm 3$) are unlikely to make up more than
$\sim$10\% of the bright SMG population and that almost all of the
$S_{\rm 1mm}\ge\rm 4$\,mJy SMG population is amenable to study via the
deepest current radio imaging, e.g.\ the $S_{\rm 1.4GHz}=\rm 2.7\,\mu
Jy$ r.m.s.\ imaging of the Lockman Hole field at 1046+59 by Owen et
al.\ (in preparation). A number of caveats remain, however: for one,
an AGN contribution to the radio flux density could mask a
high-redshift population.

\section{SMGs as scaled-up ULIRGs?}

It is often argued that SMGs resemble scaled-up ULIRGs, with compact,
intense starbursts occurring on scales of a few kpc or
less. Measurements of small source sizes for their radio or CO
emission are often cited to support this notion (e.g.\ Tacconi et al.\
2006). Evidence exists, however, that this compact mode of galaxy
formation is not necessarily ubiquitous amongst high-redshift
galaxies, nor would we expect it to be when we consider the Eddington
limit for $\ge$10$^3$-M$_{\odot}$\,yr$^{-1}$ starbursts. Ivison et
al.\ (2000) and Stevens et al.\ (2003, 2004) presented images of
distant, dusty AGN -- radio galaxies and X-ray-absorbed quasars --
showing rest-frame far-IR emission on scales of 10s of kpc, though
whether that emission is due to smooth, galaxy-wide events or due to a
number of intense starbursting clumps has so far been impossible to
ascertain.

\setcounter{figure}{7} \begin{figure}[!t]
\plotfiddle{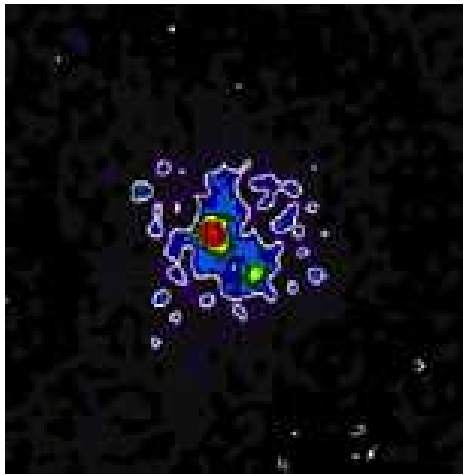}{5.8cm}{0}{160}{160}{-85}{-1028}
\caption{$\rm 5''\times 5''$ MERLIN+VLA 1.4-GHz image (0.2$''$ {\sc
fwhm}), with unprecedented sensitivity ($>$1\,Ms of integration),
of an SMG at $z=\rm 1.2$
(Biggs \& Ivison, in preparation). The radio emission, which should be
thought of as a high-resolution surrogate for the far-IR emission, is
extended (unlike the optical) on a scale of $\ge$2$''$
($\ge$15\,kpc), with several compact knots.}  \end{figure}

Bright SMGs for which high-resolution submm/mm imaging have been
sought to address this question, thus far in vain, include HzRG850.1
(Ivison et al.\ 2000). Now, however, exceptionally deep,
high-resolution radio imaging with the MERLIN array (Biggs \& Ivison,
in preparation) -- an unprecendented 1-Ms integration, combined in
the $uv$ plane with 270\,ks of VLA data using a new technique -- means
that we can investigate the scale of the radio (and thus far-IR)
emission in normal SMGs. Fig.~8 shows one example: LH\,1200.08 at
$z=\rm 1.2$. Its radio emission is extended on a scale of at
least 2$''$ ($\ge$15\,kpc), with several compact knots. As such, it
bears only a passing resemblence to low-redshift ULIRGs.

It is interesting to speculate about how this object would appear if
it were at observed with the resolution and sensitivity more usually
available to us and it lay at the median spectroscopic redshift of
SMGs? The answer: it would resemble many of the existing radio
detections -- a faint, barely-resolved radio source displaying very
little evidence of the galaxy-wide starburst apparent in this deep MERLIN
image.

\acknowledgements

Many thanks to Steve Eales, Thomas Greve, Andy Biggs, Thomas Targett,
Eiichi Egami, James Dunlop, Ian Smail, Andrew Blain, Scott Chapman,
Jason Stevens, Mat Page and Dave Alexander.

\end{document}